\documentclass[a4paper,12pt]{article}
\usepackage[utf8]{inputenc}
\usepackage{graphicx}
\usepackage{float}
\usepackage{bm}
\usepackage{hyperref}
\usepackage{authblk}
\RequirePackage[top=16mm,bottom=16mm,left=25mm,right=25mm,head=15mm,includeheadfoot]{geometry}
\begin{document}

\title{\textbf{Matter in Discrete Space-Times}}
\author[]{{{\bf P. P. Divakaran}} \\ 
{\small (Email: \href{ppdppd@gmail.com}{ppdppd@gmail.com})} }
\affil[]{{\small Riviera Retreat, Kochi, 682013 (India).}}

\date{}
\maketitle

\begin{abstract}
In the Einsteinian model of space-time as a 4-dimensional pseudo-Riemannian manifold, special relativity holds exactly in the tangent space of every point. A quantised matter field of a given mass and spin, corresponding to an elementary particle of matter, is then to be regarded as being defined by a unitary representation (UR) of the Poincar\'{e} group at each point. This Wignerian viewpoint leads to a more general reformulation of the equivalence principle as the unitary equivalence of these URs as the point is varied. In this background, the main question addressed in these notes is whether, as a necessary first step in a discretisation of gravity, the Wigner construction can be carried over to a model space-time which is a 4-dimensional lattice embeddable in real Minkowski space with a distance function inherited from it (but physically not so embedded). Working with a hypercubic lattice, it is shown in full mathematical detail that the Wigner paradigm continues to be valid but with some exotic new features. The description of spin is essentially the same. In contrast, the momentum space is the 4-torus, identified as the Brillouin zone  of  space-time where all physical phenomena occur: 4-momentum is defined and conserved only modulo a reciprocal lattice vector, implying that there is no notion of an invariant mass except when it vanishes. Nevertheless, massless particles continue to have a constant invariant speed (the `speed of light'), a result of crucial importance for the viability of discrete relativity. A massive particle in contrast is characterised by a rest mass and under large boosts it will pass through phases of superluminal propagation. If the lattice spacing is taken as a fixed fundamental length of the order of the Planck length, such effects can be observed only in the evolution of the very early regime of the conventional big bang universe, of which the two most dramatic manifestations are i) cosmic Umklapp processes, leading to a degradation of energies of individual particles, as a possible source of ultra-high energy cosmic rays and ii) primordial superluminal expansion as a contribution to or even as the root cause of cosmic inflation. A fundamentally discrete space-time is not in conflict with known physics; it may in fact be of help in explaining some otherwise mysterious aspects of early cosmology.

\end{abstract}
\maketitle
\section{A fundamental length?}
Explorations of the possibility that space or space-time may have a discrete structure on a scale much smaller than currently accessible have a long history. Though the motivations for considering such a fine structure have changed somewhat over time, they have a common origin in the old observation of Planck that the fundamental constants $c$ and $\hbar$, which are independent of dynamics, can be combined with the coupling strength $G_{N}$ of the gravitational interaction of matter in the Newtonian limit to produce a unit of length $L=(\hbar G_{N}/c^{3})^{1/2}$, of a far smaller scale than any length we can conceivably measure. The expectation, supported by ingenious thought experiments and some theoretical considerations, is that in a quantum theory of gravity (if and when we succeed in constructing one), or even in a semi-classical approximation to it, distances of the order of $L$ or smaller cannot have an operational significance.\footnote{There is an extensive review of work in this area, particularly valuable for the historical and motivational background, in [1]. It is also very useful for its comprehensive bibliography as of the time of its writing. New work continues to be produced in profusion but it seems fair to say that there has been no significant recent breakthrough.}

We may then consider two physically distinct types of models of space-time as posssible ways of accommodating a fundamental length. The more popular is one in which space-time is still a pseudo-Riemannian manifold but does not admit physical measurements of lengths below a finite length much smaller than any that is accessible to present day methods (which we may take, at least tentatively, to be the Planck length $L$). Models of this type have problems of reconciliation with principles we hold to be inviolable, such as the observer-dependence of the magnitude of $L$, making its interpretation as a fundamental characteristic of space-time itself somewhat questionable. There are many variants of such models, not always mutually compatible, but all of them have to invoke some sort of deformation of special relativity, whose consequences are not fully understood. 

A more ambitious approach is to take space-time to be fundamentally discrete, not embedded physically in a manifold (but of course, mathematically, so embeddable). The fundamental dynamical variables -- the counterparts of local fields on manifolds -- are then to be defined on the points of a discrete set (and only on them because that is all there is) endowed with a notion of ordering, a lattice, derived perhaps from the metric in an embedding manifold. It will obviously be a formidable undertaking, already at the classical level, to transcribe the geometry of general relativity to lattice space-times. Different types of lattices have been studied in the literature and it is fair to say that it is too early for a prognosis of where these studies might lead, see the review [1] and the many references therein. To then extend it to the quantum domain will be an even greater challenge. An idea of the conceptual and technical difficulties that have to be overcome may be had from, for example, the reviews [2] and [3] (written some years ago; the situation is much the same today). 

There are very good reasons, nevertheless, for hoping that the successful incorporation of a fundamental length in physics at very short distance (reciprocally, very large momentum) scales can be a major step forward in an eventual quantisation of gravity. First, it is the gravitational constant $G_{N}$ itself that allows the introduction of a natural lattice structure for space-time through $L$. Indeed, in a world without gravitation, the Planck length vanishes; conversely and more speculatively, if $L$ is taken to 0 keeping $c$ and $\hbar$ finite, the space-time manifold will tend to flatness. Secondly, in a quantum or semiclassical framework, $L^{-1}$ provides a momentum and energy cutoff that makes possible the calculation of finite gravitational effects, conventionally unrenormalisable;\footnote{The first attempts at using a discrete space-time to make perturbatively finite (special) relativistic quantum field theory date from the 1930s ([1], section 2), well ahead of the time standard perturbative renormalisation theory came to be developed. Subsequently, much effort has gone into exploring the lack of renormalisability of gravitationsl interactions and its possible cures.}
And, finally, current attempts at quantising the geometry of space-time, even without a fundamental length introduced {\it ab initio}, seem to point to the need for imposing a discrete structure (as in loop gravity for instance). 

The Einstein equations not only have a left side originating in the geometry of the space-time manifold ${\cal M}$. They have also a right side concerned with matter. At the fundamental level -- the level at which ideas such as quantisation can be meaningfully addressed -- this takes the form of contributions of various matter fields to the energy-momentum density. The description of matter fields involves only flat space-time, the tangent spaces to ${\cal M}$, essentially because of the equivalence principle (as briefly recollected in the next section). Therefore any discretisation of ${\cal M}$ will entail a discretisation of flat Minkowski space $M$ in its guise as a tangent space and it becomes necessary then to study lattices in $M$ and to ask whether matter fields can be defined on them in physically satisfactory and mathematically unambiguous terms. To take a first serious look at this particular question -- a necessary step and one which is much less daunting than the discretisation of gravity -- is the purpose of this article. Although there are some mathematical issues which are yet to be fully resolved, the first answers are affirmative: `local' matter fields can be defined on regular lattices, without contradicting physics as we know it at length scales that are accessible today, but are still enormously large in comparison to $L$. As is to be expected, intriguing  new physics, with possible (and perhaps desirable) cosmological consequences, does emerge at the scale of $L$. 

Even though the discrete space-time envisaged in this paper is not to be thought of as a subset of a true physical continuum space-time, the working out of the discrete physics will often use some of the very detailed knowledge we have of the continuum physics as a model. For that reason, there will be occasional recapitulations of well-known facts from standard theory, sometimes in a more  general form than we are used to. This is especially the case in the next section on the relevance of the equivalence principle in defining matter fields and in the summary (section 5) of their construction as representations of the group of special relativity. Also, for the benefit of readers who may not be interested in the unavoidable but often tedious arguments in the main part of the paper, I have given in the next few sections a schematic outline of some of the novel points that arise.   

Throughout this paper, ${\cal M}$ and $M$ have dimension (1,3) with the metric having the signature $(+,-,-,-)$.

\section{Matter fields: tangent spaces and a general equivalence principle}
In general relativity, the bridge between the two sides of the Einstein equation is the classical equivalence principle, usually stated simply as the equality of the inertial mass and the gravitational mass. A natural starting point for the description of matter within general relativity is therefore the equivalence  principle itself,\footnote{Rewritings of the standard field equations, e.g., the Dirac equation, so as to make them consistent with GR have a long history; most of them do not explicitly invoke the equivalence principle.} 
suitably reinterpreted where necessary. In this section, I recall in a qualitative way the relevant issues. At a fundamental (`elementary particle') level, there are two steps involved. Firstly, matter fields contributing to the energy-momentum are defined not globally on ${\cal M}$ but locally at each point $x\in {\cal M}$ -- specifically, on the tangent space $T_{x}{\cal M}=:M_{x}$ which is by defintion flat and whose spatial projection is  the local inertial space at $x$. The reason of course is that an invariant mass is an attribute associated with the Poincar\'{e} group $P$ operating on flat Minkowski space. Thus a complete theoretical understanding of the principle as first formulated by Einstein had to wait till Wigner's landmark paper ([4], see also [5]), according to which an invariant mass is one of the parameters labelling   irreducible unitary representations (URs for short) of the Poincar\'{e} group $P$.\footnote{Several variants of this foundational principle, ranging from the descriptive to the abstractly mathematical, are still current in the literature, more than a century after Einstein first proposed his version of it. The specal theory is not primarily just a weak field approximation to the general theory but a constituent part of it that holds exactly in every tangent space, thereby defining precisely the intuitive notion of an inertial frame; that is the content of the equivalence principle. The formulation given here is a modern, physically and mathematically sharp version of the original Einstein formulation that takes account of some key insights that came later, such as the significance of symmetries in quantum theory.}
(Precise characterisations of this and other relevant groups will be given below as and when they are needed).
It follows that a logical precondition for the formulation of the equivalence principle is a precise notion of an equivalence of the tangent spaces as $x$ is varied, and a consequent criterion for the independence of the mass of a test particle of the point $x$ in space-time where it is measured. It is to be noted that this condition requires only the special theory for its formulation. Only after its validity is accepted can the second step, the assertion of the equality of this common inertial mass with the gravitational mass, which is otherwise a property deriving from the global geometry of ${\cal M}$, be taken. 

As everyone knows, Wigner's main result is that an irreducible UR of the universal covering group $\bar{P}$ of $P$, satisfying certain physically reasonable conditions (absence of tachyons and boundedness of helicities) is characterised by a fixed non-negative real number (the square of the mass) and a fixed set of integral or half-integral  helicities (`spin'). In other words, if we ignore `internal quantum numbers' (`charges'), such an irreducible UR of $\bar{P}$ (a Wigner UR in short) can be identified with an elementary particle with a given mass (including vanishing mass) and spin. The Wigner construction has been much studied in the literature (and will be very briefly reviewed below as it is the basis on which the description of matter in discrete space-times rests). Here I make two qualitative observations which can be expected to have a bearing on the general philosophy of matter fields in a theory of gravity already in the `continuum limit'.

1.  Wigner's construction is the foundation of a {\it quantum} description of matter. The Hilbert space ${\cal H}^{1}$ on which an irreducible UR of $\bar {P}$ is realised is the 1-particle state space. An $n$-particle state then belongs to the $n$th tensor power $\otimes ^{n}{\cal H}^{1}=:{\cal H}^{n}$ (symmetrised or antisymmetrised according to whether the particle is a boson or a fermion) and a general state in a quantum field description of the system is a linear sum of tensor product states (with decay conditions on the coefficients to ensure that they form a Hilbert space ${\cal H}$) -- creation and annihilation operators are essentially operators which map ${\cal H}^{n}$ to ${\cal H}^{n+1}$ and ${\cal H}^{n-1}$ respectively. This is standard and the point of bringing it up is to highlight the fact that in a putative quantum theory of gravity, the description of matter {\it \`{a} la} Wigner can be expected to remain valid. Indeed, we have no other way of characterising matter at the quantum level, a recognition that is implicit in all current work.

2. To give meaning to the notion of the identity of a matter particle independently of its gravitational environment -- to be able to say as is commonly done that an electron is an electron whether in a vanishingly weak gravitational field or near a black hole -- it is necessary to formulate the equivalence principle somewhat more generally. A reasonable and obvious generalisation would be to postulate that the Wigner UR $U$ corresponding to a given particle (not just its partial attribute, the mass) is abstractly the same at all points $x\in{\cal M}$. The qualification `abstractly' is necessary since, even though the group $\bar{P}_{x}$ is, abstractly, the same at all $x$, as a transformation group on the tangent space $M_{x}$ it depends on $x$ and so does the UR $U_{x}$ as the unitary group of ${\cal H}^{1}_{x}$ (where I have distinguished structures localised at $x$ by a subscript). This $x$-dependence of the representation as a whole (unlike its Casimir, the mass, which is supposed constant over ${\cal M}$), leads to an immediate natural linkage between matter fields and gravity: specifying a connection (or some other equivalent object such as the covariant derivative) on the tangent bundle of ${\cal M}$ allows the parallel transport of a frame in $M_{x}$ to $M_{y}$ (together with their induced Minkowski metric), and hence of their group of isometries: $\bar{P}_{x}$ to $\bar{P}_{y}$, and hence of a given UR: $U_{x}$ to $U_{y}$. Since these representations are abstractly identical, they will correspond to the same mass and spin. The equivalence can in fact be stated in a form respecting the quantum nature of the Wigner description of matter: there is a unitary map  $W(x,y): {\cal H}_{x}^{1}\rightarrow{\cal H}_{y}^{1}$ such that $ W(x,y)U_{x}=U_{y}W(x,y)$.

The setting best suited to the exploration of the interrelationship between gravity and matter -- equivalently, of the geometry of ${\cal M}$ and the representation theory of the isometries of $M_{x}$ -- is thus that of the tangent bundle over the space-time manifold. Its full working out will be a major undertaking. What is already clear from the above very preliminary remarks is that the matter side of the equation is not only an essential part of such a project, but may even be a good starting point in the search for a quantum theory of gravity, much as the formulation of quantum electrodynamics is most profitably approached by starting with the invariance properties of charged fields. 

In any case, independently of how such a programme will work out,  one cannot avoid dealing with the representation theory underpinning the description of matter. In the continuum case, the representations are very well understood from Wigner's work. Their significance as the foundation of a quantum definition of matter is also well known. It is less well appreciated that in that role it is central to the description of matter in the general theory as well, by enabling the original formulation of the equivalence principle to be generalised and completed as indicated above. The purpose of the present work is however more limited: to show how that description can be adapted to the case of space-time being discrete and to bring out the ways the results deviate from standard wisdom.  The most crucial of these deviations will turn out to result in a departure from causal propagation of massive matter at extremely high (Planck scale?) speeds,  perfectly acceptable  in our present state of knowledge, perhaps even desirable in current models of very early cosmology.

\section{Discrete Minkowski space and its symmetries}

The core of this paper is concerned with the question of whether and how elementary matter fields can be associated with URs of the restriction of the Poincar\'{e} group to a suitably chosen discrete subgroup. Answering the question will involve the following steps:

1. The choice of a discretisation of $M$, i.e., ${\bf R}^{4}$ with the Minkowski metric. The simplest choice is the hypercubic lattice ${\bf Z}^{4}$ of points in ${\bf R}^{4}$ with integral coordinates with respect to a fixed set of axes. This means in particular that space and time coordinates are related by the speed of light, assumed to be a fundamental constant and put equal to unity. (Or, equivalently, the speed of light is defined as the ratio of the spatial and temporal lattice spacing -- but measured in what units?). Thus there is a unique lattice spacing which is the unit of length and time, also put equal to unity in most of what follows. (Where necessary, the Planck length will be brought in to play that role). From the point of view of symmetries, this is the simplest choice; other regular lattices will not pose any conceptual problems but will add to the technical and computational burden. Random lattices are excluded from consideration since they will entail randomly distributed lattice spacings and cannot naturally accommodate a unique fundamental length.  

A distance function is defined on ${\bf Z}^{4}$ by restricting the Minkowski metric in ${\bf R}^{4}$ to its integral points: if $X=\{X_{\mu}\in{\bf Z};\mu=0,1,2,3\}$ is a point of ${\bf Z}^{4}$, its (length)$^{2}$ is given by $X_\mu X_\mu :=X_{0}^{2}-X_{1}^{2}-X_{2}^{2}-X_{3}^{2}=:X^{2}$ (and similarly for the (distance)$^{2}$ between two points $X$ and $Y$). The discrete set ${\bf Z}^{4}$ together with this distance function is our discrete Minkowski space and will be denoted by $M({\bf Z})$ in what follows.

2. The identification of the lattice Poincar\'{e} group. In the continuum, the relativity group is $P({\bf R})=L({\bf R})\vec{\times}T({\bf R})$, where $L({\bf R})=SO(3,1,{\bf R})/\{\pm 1\}$ is the connected (proper, orthochronous) Lorentz group, $T({\bf R})$ $(\sim{\bf R}^{4})$ is the translation group (in a more explicit notation than in the introductory remarks) and $\vec{\times}$ denotes the semidirect product, the arrow indicating that (the quotient group) $L$ operates on (the normal subgroup) $T$. The connectedness requirement on $SO(3,1)$ (the quotienting by $\{\pm 1\}$) keeps out reflections which are not, observationally, symmetries of matter field interactions among themselves even though they leave the metric invariant. The discrete Lorentz group is then the subgroup of $L({\bf R})$ obtained by restricting every $4\times 4$ matrix $\lambda\in L({\bf R})$ to have integral entries: $L({\bf Z}):=SO(3,1,{\bf Z})/\{\pm 1\in SO(3,1,{\bf Z})\}$.  Our discrete Poincar\'{e} group $P({\bf Z})\subset P({\bf R})$ is therefore the semidirect product of this group with the discrete translation group $T({\bf Z})$ $(\sim{\bf Z}^{4}\subset {\bf R}^{4}$). It is an interesting fact that, while $L({\bf Z})$ (generalised in the obvious manner) is the 2-element  group in 1 + 1 dimensions, it is an infinite group in all higher dimensions.

3. Determination of the appropriate representations of $P({\bf Z})$. The guiding spirit in identifying and constructing the representations will be the work of Wigner on the corresponding problem for $P({\bf R})$. Technically, this will be the major concern of the present paper. Here I limit myself to describing the physics and mathematics background to the identification of the relevant representations. The starting point is the recognition (due, also, to Wigner [6]) that the group of symmetries of a quantum system is represented on its state space by projective URs. Wigner ([4]) first establishes the result that every continuous projective UR of $P({\bf R})$ lifts to a continuous UR of its universal covering group $\bar{P}({\bf R})=\bar{L}({\bf R})\vec\times T({\bf R})$, with $\bar{L}({\bf R})=SL(2,{\bf C})$; i.e., given a projective UR of $P({\bf R})$, we can find a UR of $\bar{P}({\bf R})$ whose projection  onto the quotient group $P({\bf R})$ is the given projective UR. This key result has the following ingredients: i) though $T({\bf R})$ has nontrivial projective URs (i.e., projective URs which are not equivalent to URs of itself), they do not extend to the whole of $P({\bf R})$ as nontrivial projective URs and can be ignored; ii) though semidirect product Lie groups $G\vec{\times}A$ with $A$ abelian can in general have nontrivial projective URs which restrict to $A$ as URs, this does not happen for $P({\bf R})$ on account of the semisimplicity of $L({\bf R})$; and iii) every projective representation (not necessarily unitary) of $L({\bf R})$ lifts to a (linear) representation of its universal cover $SL(2,{\bf C})$, again because of semisimplicity. The realisation of an irreducible PUR for a particle of a given mass and a given spin, integral or half-odd-integral, then involves the choice of a mass shell (an orbit of $L({\bf R})$ in momentum space) and a finite dimensional (necessarily non-unitary) representation of $SL(2,{\bf C})$ which determines the spin.  

These results, in particular the assertion iii), are obviously specific to Lie groups and are therefore not directly applicable to the projective URs of their discrete subgroups. There are however theorems relating projective representations (not limited to projective URs) of any group to linear representations of related `universal' groups other than the universal cover. Specifically, given a group $G$, we can construct a group $\hat{G}$, called a universal central extension of $G$, of which $G$ is a quotient group, and having the property that every projective representation of $G$ is the projection of a linear representation of $\hat{G}$.\footnote{Even for Lie groups, $\hat{G}$ is not necessarily $\bar{G}$. Questions regarding projective representations of $G$ are most efficiently addressed in the language of the theory of central extensions of $G$ by appropriate abelian groups and the associated group cohomology theory. For a clear and thorough treatment of the topic, including the construction of universal central extensions, see Raghunathan ([7]) and, for a physicists' version with applications to many examples, see Divakaran ([8]). The theory can in fact be used to reformulate the process of quantisation of a system entirely in terms of its symmetries in a manner free from commonly encountered ambiguities ([9]); in particular, the superselection structure of the state space is seen to be of cohomological origin, a fact which may play a role in the specification of particle states in discrete relativity (see section 7 below). Wigner's work ([4]) was of course the first to determine the state space of an elementary particle explicitly as a projective UR of $P({\bf R})$ but he did not connect it to the consequent superselection rule, that of univalence.} 
The statements i) to iii) above are in fact specialisations of properties of universal central extensions to Lie groups having different structural properties. In particular, iii) follows from a theorem which says that a connected semisimple Lie group has a unique universal central extension and that it is the same as its universal cover; so equivalence classes of its projective representations are classified by the Pontryagin dual (the group of 1-dimensional representations or characters) of its fundamental group. This is the reason why it is legitimate to work with $\hat{L}({\bf R})=\bar{L}({\bf R})=SL(2,{\bf C})$.\footnote{Many commonly met groups in physics serve as examples of the distinction between $\hat{G}$ and $\bar{G}$. Thus, the universal cover of the 2-dimensional rotation group $SO(2)$ is the real line but its universal extension is itself (it has no nontrivial projective UR -- hence no non-integral spin) while the vector (translation) group ${\bf R}^{n},\,n>1$ has nontrivial projective URs but is its own universal cover. The indiscriminate substitution of $G$  by $\bar{G}$, rather than by the always correct $\hat{G} $ has led to much misunderstanding in the physics literature, see [8].} 
 
To deal with the discrete groups of interest to us with anything like this degree of completeness is not a feasible option, primarily because of the lack of a physically satisfactory criterion (such as continuity) for acceptable representations. And it is, to a great extent, unnecessary for our purpose; physically, it is sufficient to note first that the discrete groups of our interest are subgroups of the corresponding Lie groups by construction and then to find those representations which, in the limit, approach in a well-defined sense the physically acceptable representations of the embedding Lie groups. That is possible thanks to the fact that $L({\bf Z})$ and $\hat{L}({\bf Z})$, as subgroups of the embedding Lie groups, have a property known as Borel density which enables them to inherit several useful results regarding their representations from the Lie groups (see below for details). This is in fact one of the mathematical inputs that make our project at all feasible. 

4. Interpreting representations as particles. The implementation of the programme outlined above presents some (though surprisingly few) serious mathematical obstacles. The resulting physical picture too, naturally, differs in some significant respects from the continuum theory. Firstly, since the momentum space (the space of characters of the translation group) is now the 4-torus ${\bf T}^{4}$ rather than ${\bf R}^{4}$, momentum itself is defined and conserved only modulo a reciprocal lattice vector (which is the same as the momentum cut-off, the Planck momentum by choice). This has consequences somewhat like the familiar momentum space properties of an electron moving in a crystal; in particular, the mass shell is the Minkowski metric analogue of the Fermi surface of an empty lattice. Secondly, the spin of a representation can no longer be defined generally as an attribute of rotation invariance as the discrete `rotation' group, being a discrete subgroup of the compact Lie group $SO(3,{\bf R})$, is a finite group with a finite set of inequivalent irreducible URs. But this is not a serious handicap since it turns out (essentially because of the Borel-density property of the discrete $SL(2)$) that spin, both integral and half-integral, can be defined by reference to the Lorentz group alone (in the continuum, the two ways of defining spin are of course equivalent).\footnote{Another respect in which the identification of representations with particle states in the discrete world differs from continuum relativity is that they may apparently be chosen to be (highly) reducible. Whether this freedom is physically significant is at present unclear, see the discussion in section 7 below.}    
If we accept these deviations from the received wisdom of continuum relativity -- which, we shall see, are not in contradiction with our current state of knowledge -- projective URs of $P({\bf Z})$, of a certain general type, have a perfectly reasonable interpretation as elementary particles.  

Of the deviations from standard lore, the more dramatic are those having their origin in the compactness of the momentum space. These include in particular possible apparent violations of energy-momentum conservation in elementary processes -- the analogue of the Umklapp processes of crystal physics -- involving energies of the order of the Planck mass. More intriguingly, the distinction between time-like and space-like momenta is no more an invariant concept. `Massive' orbits of the discrete Lorentz group in momentum space do not have an invariant mass associated to them and have tachyonic branches that begin to sprout around the Planck scale: the (energy-momentum)$^{2}$ can be negative even when the (rest-mass)$^{2}$ (which, being a zero-momentum attribute, is a valid concept) is positive. The light cone itself is well-defined as a closed hypersurface in the momentum space ${\bf T}^{4}$; zero mass orbits lie within it and have no tachyonic branches. (This is a result of independent and fundamental importance, as described in a separate added note.) The general scheme for the construction of URs following from these considerations will be described later on with a degree of mathematical detail, as well as, qualitaively, some of the unfamiliar physical consequences of discrete relativity. What is certain is that the exotic features that emerge have no impact on elementary particle phenomena at energies presently accessible to experiments or many orders of magnitude higher; they will, however, have cosmological implications which also will be touched upon at the end.

The main results of this work, then, hold no bad surprises: the description of elementary matter as founded on special relativity survives discretisation, subject to some reinterpretations which are capable of being tested. But these are only the first steps and much still needs to be done. Moreover, beyond the continued validity of the Wigner definition of elementary particles, a lattice structure for space-time as put forward here will have other macroscopic manifestations, `macroscopic' in the present context meaning (here as elsewhere in this article) length scales characterising the structure and interactions of the currently known particles and greater: issues such as possible deviations from isotropy, the accommodation of truly macroscopic (in dimension and/or mass) classical systems etc. They have been much written about in the literature (see [1] and the many references therein) and this paper will have not much to add to it.

These introductory sections are meant to bring out, more or less qualitatively, the following points: a) the critical importance of a good description of matter fields -- including a suitable formulation of the equivalence principle -- before we can think of bringing together quantum mechanics and general relativity; b) the feasibility of such a programme in a discrete space-time; and c) a brief foretaste of the necessary (but not widely known in the physics literature) mathematical material, described in the next few sections, that it entails.

\section{The discrete Lorentz and Poincar\'{e} groups and their central extensions}

In the rest of this paper, the connected real Lorentz group $L({\bf R})$ will be called simply the Lorentz group without any qualifiers and its discrete subgroup $L({\bf Z})$ the discrete Lorentz group (and corresondingly for the Poincar\'{e} groups). As noted in section 3, the semisimplicity of the Lorentz group has the consequence that the universal central extensions of $L({\bf R})$ and $P({\bf R})$ are in fact their universal covering groups and hence that all their projective representations can be obtained as projections of true (linear) representations of the universal covers. To repeat, this is the reason why $L({\bf R})$ is replaced by $\bar{L}({\bf R})=SL(2,{\bf C})$ in the determination of physical (i.e., projective) URs of the full symmetry group of special relativity and, hence, for their interpretation as the state spaces of elementary particles.   

Lacking the completeness of such Lie-theoretic concepts and results, the method followed here has the limited aim of finding certain finite-dimensional projective representations of $L({\bf Z})$ (and, eventually, of projective URs of $P({\bf Z})$) which are inherited naturally from those of $L({\bf R})$. In other words, we shall look for a certain subgroup $\hat{L}({\bf Z})$ of $SL(2,{\bf C})$ having the property that every projective representation of $L({\bf Z})$ that is the restriction of a projective representation of $L({\bf R})$ lifts to a linear representation of $\hat{L}({\bf Z})$. This requirement is met if $\hat{L}({\bf Z})$ has a central ${\bf Z}_{2}$ subgroup such that $\hat{L}({\bf Z})/{\bf Z}_{2}=L({\bf Z})$, exactly as in the corresponding Lie group case where it is a standard construction found in text books (see for example [10]). In fact, the result in the discrete case is a direct transcription of this standard construction which I therefore recall.  

Denote by $H(2,{\bf C})$ the real vector space of $2\times 2$ complex hermitian matrices and by $\tau_{i}, i=1,2,3$, the Pauli spin matrices. The association of $x\in M$ to the matrix $\mathrm{x}:=x_{\mu}\tau_{\mu}$ ($\tau_{0}=$ unit matrix), with $x_{\mu}=(1/2)\mathrm{tr}(\tau_{\mu}\mathrm{x})$, is a bijection of $M$ and $H(2,{\bf C})$, such that $x^{2}=\mathrm{det(x)}$. $SL(2,{\bf C})$ has an action on $H(2,{\bf C})$ preserving the determinant: $\mathrm{x}\rightarrow \alpha\mathrm{x}\alpha^{*}, \alpha\in SL(2,{\bf C})$. Correspondingly, for any $\alpha$ there is a $\lambda\in L({\bf R})$ such that $(\lambda x)_\mu \tau_\mu =\alpha\mathrm{x}\alpha^*$ and hence a homomorphism $SL(2,{\bf C})\rightarrow L({\bf R})$ whose kernel is easily seen to be the central subgroup ${\bf Z}_{2}=\{\pm 1\in SL(2,{\bf C})\}$.  

Given this explicit identification of $SL(2,\bf C)$ as the (unique) nontrivial central extension of $L(\bf R)$ by $\bf Z_2$, its adaptation to the discrete case is straightforward, thanks to the fact that the basis $\{\tau_\mu\}$ of $H(2,\bf C)$ are actually matrices over the ring ${\bf Z} [i]$ of Gaussian integers, i.e., complex numbers whose real and imaginary parts are integers. Replacing $M$ by its discrete counterpart $M({\bf Z})$ therefore gives a bijection of $M({\bf Z})$ and $H(2,{\bf Z} [i])$ exactly as in the real case (even though they are no longer vector spaces), $X\rightarrow X_\mu\tau_\mu=:\mathrm{X}$, such that $X^2=\mathrm{det(X)}$. And, as before, i) the group $SL(2)$ over Gaussian integers, $SL(2,{\bf Z} [i])\subset SL(2,\bf C )$, acts on $H(2,{\bf Z} [i])$ by $\mathrm{X}\rightarrow A\mathrm{X}A^*$; ii) given any $A\in SL(2,{\bf Z}[i])$, there is a discrete Lorentz transformation $\Lambda$ such that  $(\Lambda X)_\mu\tau_\mu=A\mathrm{X} A^*$ and a homomorphism $SL(2,{\bf Z}[i])\rightarrow L({\bf Z})$; and iii) the kernel of this homomorphism is ${\bf Z}_2 =\{\pm1\in SL(2,{\bf Z}[i]\}$.  In other words, $ SL(2,{\bf Z}[i])$ is a nontrivial central extension of $L({\bf Z})$ by ${\bf Z}_2: SL(2, {\bf Z}[i])/{\bf Z}_2 =L({\bf Z})$; every finite dimensional representation of $SL(2,{\bf Z}[i])$ will project to $L({\bf Z})$ as a projective representation, either as a true representation or as a `representation up to sign'. This is the reason why this group is denoted by $\hat{L}({\bf Z})$; it plays the same role in the theory of projective representations in discrete relativity  as $\hat{L}({\bf R})=SL(2,\bf C)$ does traditionally. In particular, it will allow for the presence of states of half-integral helicities in discrete quantum relativity as we shall see below.

As noted in section 3 above, the full discrete relativity group -- the discrete Poincar\'{e} group --  is the semidirect product group $P({\bf Z})=L({\bf Z})\vec{\times}T({\bf Z})$, $T({\bf Z})\sim {\bf Z}^4$ being the discrete translation group; it is this group whose projective URs we would like to interpret as the state spaces of elementary matter. By the general theory of central extensions, they can be obtained as the (linear) URs of the corresponding central extension\footnote{In general, inequivalent central extensions of any group $G$ are classified by the second cohomology group $H^2(G)$ with appropriate coefficients. $H^2$ of groups having a semidirect product structure $G\vec{\times}A$, $A$ abelian, can have contributions other than just $H^2(G)$. They are absent in the continuum case (see section 3) and, for that reason, will be ignored (if they are present; the answer is not known to me) in the discrete case as well. See, however, the discussion in section 7 on the role of irreducibility in the particle interpretation of URs of $\hat{P} ({\bf Z})$.

For completeness I add that, in the semidirect product, the action of $\hat{L}({\bf Z} )$ on $T({\bf Z})$ is the same as that of $L({\bf Z})$, extended by letting the central ${\bf Z}_2$ act trivially, as in the real case.} 
 $\hat{P}({\bf Z})=SL(2,{\bf Z}[i])\vec{\times}{\bf Z}^4$.   
  
\section{Unitary representations of $\hat{P}(\bf R)$ -- an overview}
For the construction of physically acceptable URs of $\hat{P}({\bf Z})$ the model will be Wigner's method of `inducing from little groups' for the corresponding Lie group [4].\footnote{Wigner's pioneering work was given a general treatment, in particular as it applies to semidirect product groups, by Mackey ([11]). There are many subsequent accounts of the method in the literature. For the positive mass URs the summary given here is based on the description in [8] and, for the massless URs, the treatment below of the group-theoretic origin of the subsidiary condition appears to be new.} 
The generality of the method allows room for its adaptation, with suitable adjustments, to the discrete case. More importantly, the method naturally highlights the physical attributes, mass and (Lorentz) helicity, that allow a direct association of elementary quantum fields with URs, thereby serving as a model for deciding which URs of $\hat{P}({\bf Z})$ can be considered `physical'. The following is a compressed description of the essential elements of the Wigner construction in the continuum case.

The momentum space is the dual group of the translation group $T(\bf R)$, isomorphic also to ${\bf R}^4$ and denoted by $M^*$ with coordinates $\{p_\mu \}$. Let $O$ be an orbit of $\hat{L}(\bf R)$ in $M^*$ for the natural action of $L(\bf R)$, lifted to $\hat{L}(\bf R)$ by letting the central ${\bf Z}_2$ act trivially,\footnote{The action of $SL(2,\bf C )$ on $M^*$ will often be denoted as $p\rightarrow \alpha p$ (as though $\alpha \in L(\bf R )$) and likewise in the corresponding discrete case. No misunderstanding is likely to arise.} 
and let $S$ be the little group (stabiliser) of any point in $O$. $O$ can then be identified with $\hat{L}({\bf R})/S$. Suppose given a finite dimensional representation $\rho$ of $\hat{L}(\bf R)$ on a Hilbert space $V$ with the property that the restriction of $\rho$ to $S$ is unitary. Denote by $\pi$ the projection of $\hat{L}(\bf R)$ onto $O$ and by $\sigma$ a section of $\pi$, i.e., a map $O\rightarrow \hat{L}(\bf R)$ such that $\pi(\sigma (p))=p$ for all $p\in O$. Finally, let ${\cal H}_{O,V}$ be the space of vector valued functions $\phi,\psi:O\rightarrow V$, square-integrable with respect to the (positive) $\hat{L}(\bf R)$-invariant measure $\omega$ on $O$.   

On ${\cal H}_{O,V}$, define a bracket $\langle \;, \;\rangle$ by
$$   \langle\phi,\psi\rangle=\int_O d\omega(p)\langle(\rho(\sigma(p)^{-1})\phi(p),(\rho(\sigma(p)^{-1})\psi(p)\rangle_V,           $$
$\langle \;,\;\rangle_{V}$ being the scalar product on $V$. If $\sigma$ and $\sigma'$ are two sections of $\pi$, it follows from $\pi(\sigma(p))=\pi(\sigma'(p))\;(=p)$ that $\sigma(p)^{-1}\sigma'(p)$ is in $S$. And, since $\rho$ restricts to $S$ unitarily, the bracket $\langle \;,\;\rangle$ is independent of the section used to define it, making it a scalar product. Moreover, it follows from the positivity of the measure that the norm of $\phi$ vanishes if and only if $\phi=0$ identically, making (the completion of) ${\cal H}_{O,V}$ a Hilbert space.

With these notions in place, one verifies first that the action of $\hat{P}(\bf R)$ on ${\cal H}_{O,V}$ given by 
$$ (U(\alpha,x)\phi)(p):=\chi_p (x)\rho(\alpha)\phi(\alpha^{-1}p), \hspace{.5cm} \alpha\in \hat{L} ({\bf R}), x\in T({\bf R}),    $$
where $\chi_p (x)=\exp (ip_ \mu x_ \mu$) is the character of $T(\bf R )$ corresponding to $p$, is a representation. It is in fact unitary because: i) we have
$$  \|U(\alpha,x)\phi\|^{2}=\int d\omega(p)\|\rho(\sigma(\alpha p)^{-1} )\rho(\alpha)\phi(p)\|^{2}_{V}  $$
using the invariance of the measure under $p\rightarrow \alpha p$; ii) $\sigma(\alpha p)$ and $\alpha\sigma(p)$ have the same projection onto $O$ and hence differ by an element of $S$, implying $\rho(\sigma(\alpha p)^{-1})\rho(\alpha)=\rho(s)\rho(\sigma(p)^{-1})$ for some $s\in S$; and iii) $\rho$ is unitary on $S$. It is irreducible whenever $V$ is irreducible under $\hat{L}(\bf R )$. In the language of induced representations, it is the UR induced by the (unitary) restriction of $\rho$ to $S\subset \hat{L}(\bf R )$. Physically, it is helpful to refer to it as the UR supported on a given mass shell (the orbit $O_m$) and ranging over a given spectrum of helicities (the representation $V$). 
 
When $O$ is a positive (mass)$^2 $ positive (or negative) energy mass shell of mass $m$ -- i.e., the orbit $O_m $ through any point $p$ with $p^2 = m^2 $, in particular through $(m,0, 0, 0)$ -- the stabiliser $S_{m} $ is isomorphic to $SU(2)$ and every irreducible representation of $\hat{L} (\bf R )$ on a Hilbert space ${\cal H}_m $ constructed as above (dropping the subscript $V$) restricts to $S_m $ as an irreducible UR. Hence the helicity spectrum of $U_m $ is determined equivalentely and alternatively by the $\hat{L} (\bf R )$ or $S_m (=SU(2))$ transformation properties of the functions $\phi$. The spin determines the helicity; in particular the number of distinct helicity states in $U_m $ is $\dim V$. Moreover, the condition that $S_m $ fixes $p$ translates as the condition 
$$  (U_m (s,x)\phi)(p)=\chi_p (x)\rho(s)\phi (p)    $$
for all $s\in S_m$. This, or rather its Lie algebra version, is the `invariant wave equation' or the free field equation corresponding to the UR $U_m$ of $\hat{P} ({\bf R})$ ([5]). In the discrete case, it is (the  discrete form of) this condition which will replace the wave equation. 

Since all known elementary particles have non-negative (mass)$^2$ and finite sets of helicities, it is customary to reject URs of $\hat{P}(\bf R )$ not having these two properties as unphysical. As will be seen below, in the discrete context there is no `invariant mass' -- a fact that has to be physically interpreted with care -- though the notion of a rest mass $m$ still makes sense; a physical UR will then have to be characterised as one supported on an orbit passing through $(m,0,0,0)$ with $m^2 \geq 0$ and ranging over a finite dimensional representation of $\hat{L} (\bf Z )$. The condition on the helicity spectrum is a powerful one already at the continuum level: in the cases where $S$ is a non-compact (Lie) group, it puts strong restrictions on its admissible URs from which the induction process may be initiated (as also will be seen below). 

Apart from the one-point orbit that is the origin of the momentum space $M^*$, there are two nontrivial $m=0$  orbits, the open upper and lower half light cones in $M^*$. To construct physical URs of $\hat{P} (\bf R )$ supported on the upper half light cone $C_ +$ for example, consider the stabiliser $S_0 $ of the representative point $(p_0 ,0,0,p_0 ), \; p_0  >0$, consisting of the upper triangular matrices of $SL(2,\bf C )$ which we may parametrise as 
$$ 
s(\theta,z)=
\left (\begin{array}{cc}
 \exp(i\theta) & z\exp(-i\theta) \\
 0 & \exp(-i\theta)
\end{array}\right), \; 0\leq \theta<2\pi, \; z\in {\bf C}. 
$$
The group law in $S_0$: $s(\theta_1 ,z_1 )s(\theta_2 ,z_2 )=s(\theta_1 +\theta_2 \; \mathrm{mod} \, 2\pi, z_1 +z_2 \exp(2i\theta_1 ))$, identifies it as the Euclidean group in 2 dimensions $E(2,{\bf R} )= SO(2,{\bf R}) \vec{\times}{\bf R} ^{2}$, with ${\bf R} ^{2}=\{(\mathrm{Re}\,z,\mathrm{Im}\, z)\}$  
on which the $SO(2)$ subgroup acts as the 2-fold cover of the circle group; physically this $SO(2)$ is in fact the group of rotations about the 3rd axis\footnote{That it covers the circle twice is the reason why massless particles can have half-integral helicities. The Lie algebra of our $E(2)$ has $J_3 ,J_1 +K_2 ,J_2 -K_1 $ as a basis, the $J$s and the $K$s being generators of rotations and boosts in the standard physics terminology. Obviously, the ${\bf R}^2$ subgroup does not relate to physical translations.} 
(more generally the direction of the momentum vector). Its characters in a (massless) UR of $\hat{P} ({\bf R})$ constitute what is generally called its helicity spectrum. In contrast to the massive case, this (Lorentz) helicity cannot be defined through the full rotation group, which is natural since a massless state cannot be transformed to rest. 

Now, a finite dimensional UR of $E(2,\bf R )$ is necessarily non-faithful; indeed the only such URs are characters of $SO(2,\bf R )$ and have the normal subgroup ${\bf R}^2 $ as kernel.\footnote{A thorough account of the unitary representation theory of $E(2,\bf R )$ is available in [12].} 
Hence $\hat{L}(\bf R )$, being simple, cannot have any finite dimensional representation restricting to $E(2,\bf R )$ unitarily. This in turn means that the procedure of inducing from the stabiliser no longer works as directly as in the massive case and has to be modified suitably. The well known way to do this ([12], [5]) is to specialise the space ${\cal H}_{0,V} $ of functions $\phi:C_+ \rightarrow V$ as defined earlier to the subspace ${\cal H}'_0$ (dropping again the subscript $V$) on which the action of ${\bf {R}}^2 (\subset E(2,{\bf R} ) \subset SL(2,\bf C ))$ is trivial: 
$$   \rho (r)\phi (r^{-1}p)=\phi (p),  \: r\in {\bf R} ^2.            $$ 
The Lie algebra form of this condition encodes the familiar subsidiary conditions satisfied by massless fields. A character of $SO(2,{\bf R} ) = E(2,{\bf R} )/{\bf {R}}^2: \rho(\theta)\phi (p)=\exp (im\theta /2)\phi(p)$ with $\theta/2\in SO(2,{\bf R} ), m\in \bf Z $, then induces a UR of $\hat{P}(\bf R )$ on ${\cal H}'_0$. The fact to be noted, especially relevant in the context of discrete relativity, is that massless finite helicity URs exist because the appropriate stabiliser has a normal subgroup with compact quotient group. It also follows that a physically acceptable irreducible UR has precisely one (Lorentz) helicity, namely the character of $SO(2,{\bf R})$ to which $\rho$ restricts.

As for representations supported on orbits with (mass)$^2 <0$, they are doubly unphysical. Not only will they correspond to particles which are tachyonic at all momenta, they will suffer from unphysical helicities as well: the stabiliser, which is $SL(2,\bf R )$, has no finite dimensional nontrivial URs at all. 

This summary of Wignerism is meant also to remind us that the general theory of quantum fields having a particle interpretation is no more than the representation theory of the group of relativistic symmetries, subject to certain physical criteria. The significance of this foundational construction as the prelude to any attempt to quantise gravity has been noted in sections 1 and 2 above. The question now is whether this identification of particles and representations can be carried over realistically to space-times which are discrete.

\section{Masses and helicities in discrete relativity}
We turn first to an examination of how the fundamental notions of mass and helicity survive the reduction of the group of symmetries from $P (\bf R )$ to $P (\bf Z )$, both for their intrinsic significance and as preparation for the construction of physically acceptable projective URs of the latter. 

The momentum space of discrete space-time is the dual group of the discrete translation group $T(\bf Z )$ ($\sim {\bf Z}^4 $), namely the 4-torus, denoted by $B$ from now on. As in other familiar lattice problems, it is useful to think of $B=M^*/{\bf Z}^ 4$ as the fundamental domain for the action of $\bf Z ^4$ on $M^*$ where $M^*$ $(\sim {\bf R}^4)$ as earlier is the momentum space of the continuum translation group and ${\bf Z}^4$ is to be identified with the reciprocal lattice. In terms of coordinates $\{p_\mu \}$ in $M^*$, $B$ is thus the cube $\{-\pi < P_\mu \leq\pi, \; P_\mu :=p_\mu \; \mathrm{mod}\; 2\pi\}$, i.e., the unit cell of the reciprocal lattice, the relativistic analogue of the Brillouin zone or, in its cosmological role, the Brillouin zone of the universe (called simply the Brillouin zone from now on). Physically, this means of course that the momentum components are defined and conserved modulo $2\pi$, a circumstance that has far more fundamental consequences here  than in the crystal physics context (keeping in mind that it is only a minuscule part of the interior of this space-time Brillouin zone, far from its boundary, that terrestrial experiments and observations and the theories that deal with them can explore). Nevertheless, as in crystal physics, we can study the action of the discrete group $\hat{L}(\bf Z )$ on $B$ by starting with its action on $M^*$ and then translating the coordinates of the image of a point in $B$ considered as a point of $M^*$ back to $B$ by some integral multiples of $2\pi$. Under this projection $ M^*\rightarrow B $, $P_\mu=\pi$ and $P_\mu=-\pi$ get identified for each $\mu$.   

Thus a generic orbit $O_B ({\bf Z})$ of $\hat{L} ({\bf Z})$ in $B$, through a given point $P\neq 0$, can be determined by first finding the orbit $O(\bf Z )$ through $P$ of $\hat{L} ({\bf Z})\subset \hat{L} (\bf R)$ in all of $M^*$ and then translating the points of $O({\bf Z})$ outside $B$ back to $B$. It is to be expected that, generically,  the orbits will be quite `wild'.\footnote{The study of the action of discrete subgroups of Lie groups on manifolds is an active field of research and is potentially of great use in the physics of discrete systems in general. }
$O({\bf Z})$ being a subset of the orbit of $\hat{L} (\bf R)$ through $P$ considered as a point of $M^*$, a good starting point then is the projection onto $B$ of the relevant orbits in the standard continuum picture. 

Consider first the orbit $O_m $ of $\hat{L} (\bf R )$ through $(m,0,0,0), \;m\neq0$, i.e., the set of points $\{p_\mu \}$ with $p_\mu p_\mu =m^2 $ (the positive energy mass shell of rest mass $m$). It projects on to $B $  as the set of points $\{P_\mu=p_\mu$ mod $2\pi\}$, the `torus mass shell' of rest mass $m$. Figure 1 is a depiction of this (for a value of $m$ chosen to be very large so as to bring out its complicated structure) as its intersection with, say, the (0,1) plane. The corresponding orbit $O_{B,m} (\bf Z )$ of $\hat{L} (\bf Z )$ for any rest mass $0<m<\pi$, our objects of interest, will be discrete subsets of such torus mass shells. 


\begin{figure}[htb]
 \centering
 \includegraphics[width=0.8 \textwidth]{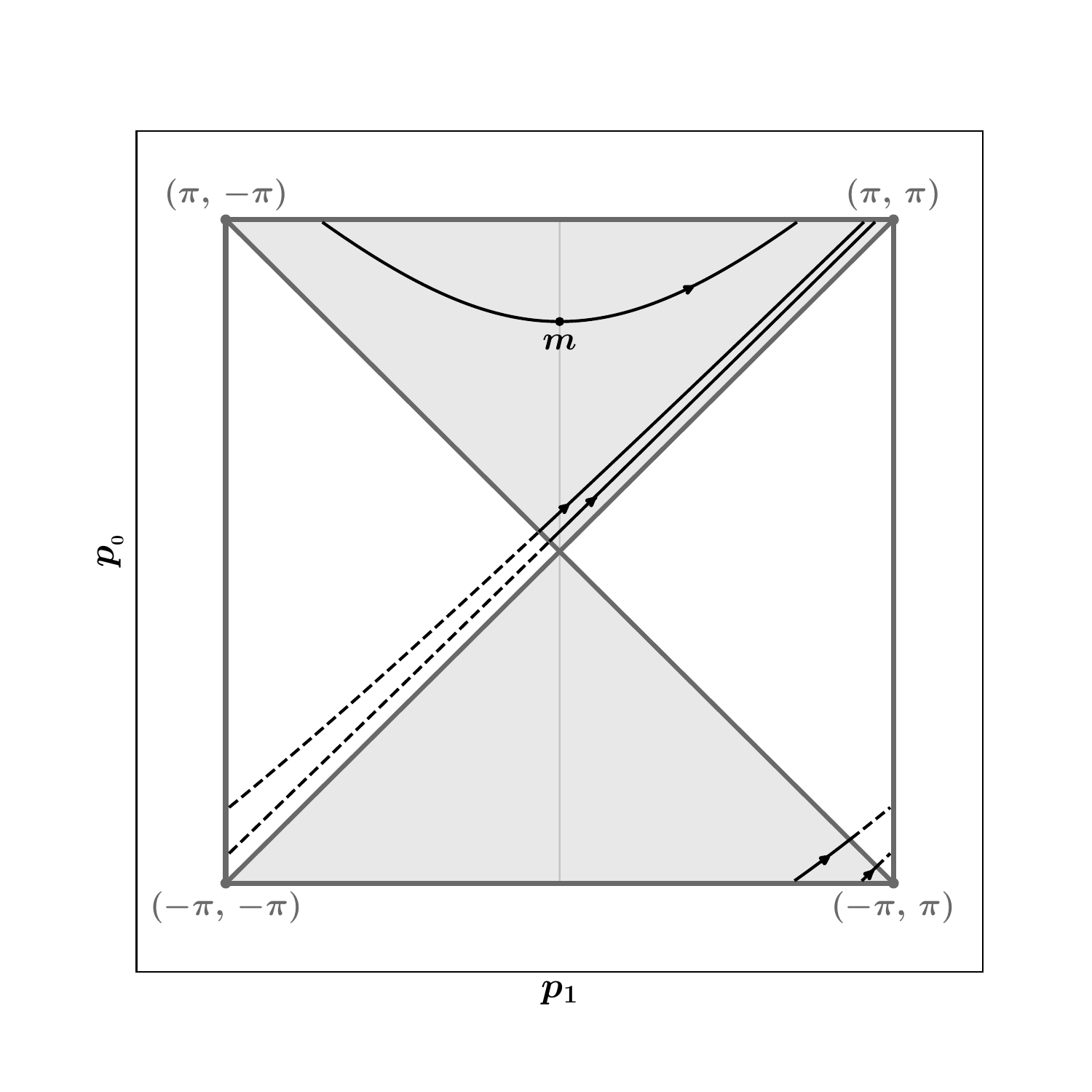}
 \caption{The intersection of a torus mass shell with the (0,1) plane. The arrows indicate that the particle is being boosted from rest in the positive 1 direction. The dotted segments illustrate the first few superluminal phases.}
\end{figure}

At the classical level, the torus mass shell summarises the kinematics -- the relationships among energy, momentum and velocity -- of a particle of a given rest mass moving in discrete space-time. The following qualitative remarks are self-evident. 

Within the Brillouin zone, i.e., where $P_\mu =p_\mu$, kinematics is entirely as in continuum relativity; in particular, the rest mass $m$ is the invariant mass for transformations in $M^*$ which keep the particle within the Brillouin zone.  But when a particle initially at rest is subjected to larger and larger boosts, say along the positive direction 1, there occur a sequence of critical points on the boundary of $B$ at which the energy or the momentum appears to undergo a discontinuous change of magnitude $2\pi$ in our units (or $2\pi /L$ in terms of the Planck length\footnote{For numerical guidance we may take it to be of the order of the Planck mass, $10^{20}$ GeV or so.}), 
induced by the identification of the points $p_\mu$ and $p_\mu + 2\pi n_\mu$ in $M^*$ for arbirary integers $n_\mu$. The first critical point occurs at $(P_0 =\pi, P_1 =\sqrt{\pi^2 -m^2 }$). A further boost in the same direction will give rise to the segment of a hyperbola (restricting the torus mass shell to the (0,1) plane as in the figure) connecting the translate $(-\pi, \sqrt{\pi^2 -m^2})$ of this point with the next critical point, namely the translate $(\sqrt{\pi^2 +m^2} -2\pi,\pi)$ of the point $(\sqrt{\pi^2 +m^2},\pi)$ (which is outside $B$). And so on {\it ad infinitum}. In all the segments except the initial one, $P_0^2 -P_1^2\neq m^2$. More generally, $P_\mu P_\mu$ is not invariant under the action of the Lorentz group, continuous or discrete; `mass' in the term `torus mass shell' refers to the rest mass as the parameter labelling distinct mass shells. This is just a reflection of the fact that there are no $\hat{L} (\bf Z )$ (and hence no $\hat{L} (\bf R )$) invariant non-zero periodic functions on $M^*$ and stems from the periodicity of the momentum components in $M^*$. 

It is also clear, most simply and graphically from Figure 1,  that all segments of the mass hyperbola except the first have parts with $P_0^2 -P_1^2 $ negative, where the particle's speed is greater than 1: under large boosts the particle behaves intermittently as a tachyon with an imaginary effective mass $m_\mathrm{eff} =\sqrt{P_\mu P_\mu }$. In the limit of an infinitely large boost, the effective mass tends to zero and the mass shell for any non-zero rest mass tends to the torus light cone. The full torus mass shell embedding a massive orbit thus has a complicated structure.To be noted in particular is that the speed of light is not an impassable barrier: a particle in a tachyonic kinematic regime can always be deboosted to rest. 
 
For ease of reference, I will refer to the part of a torus mass shell on which the rest mass is also the invariant mass as the `conventional part of the mass shell' or `the conventional regime'. It consists of the orbit through $(m, 0, 0, 0)$ before it hits the first critical point $(P_0=p_0=\pi,|\vec{P}|=|\vec{p}|=\sqrt{\pi^2-m^2})$; none of the exotic aspects of the new kinematics manifest themselves when the energy and momentum are constrained to be in the conventional part of a mass shell.

In contrast, consider the light cone $C$ of $M^*$. Under projection onto $B$ (translations by multiples of $2\pi$), its image $C_B$ remains in the torus light cone, the part of the light cone lying in $B$ (with the usual boundary conditions), In consequence, any boost of a point in $C_B$ will take it first to a point in $C$ and then, on translation back into $B$, to a point in $C_B$ itself. Masslessness is an invariant property in discrete relativity -- the polynomial $p_\mu p_\mu$ is periodic and invariant as long as it vanishes. The orbit $O_{B,0}$ of any point in $C_B$ for the action of $\hat{L} (\bf Z )$ on $B$ is a set of discrete points in $C_B$. 

As will be seen below, this is a property of capital importance.

Next is the problem of whether and how helicities can be defined in discrete space-time. They cannot be defined as arising from the URs of the rotation group by restriction since $SO(3,\bf R )$ or $SU(2)$, being compact, has only finite groups as discrete subgroups and therefore cannot accommodate general spins. The way out is to define helicity relativistically, as arising directly from the Lorentz group, without invoking the rotation group and without transforming to rest. (Recall the discussion in section 5 of how massless helicities are defined in the Wigner construction). We have in fact the key result:

Every finite dimensional irreducible representation of $\hat{L} (\bf R )$ restricts to its discrete subgroup $\hat{L} (\bf Z )$ as an irreducible representation.

This is a special case of a general theorem, the density theorem of A. Borel ([13]; see also [14] for an account of the general theoretical framework), on representations of discrete subgroups of non-compact semisimple Lie groups. A general formulation of the theorem is as follows. Let $G$ be a semisimple Lie group none of whose factors is compact and $\Gamma$ a discrete subgroup of $G$ having the property that the quotient $G/\Gamma$ has finite volume (i.e., $\Gamma$ is sufficiently dense in $G$). Then every finite dimensional irreducible representation of $G$ restricts to $\Gamma$ irreducibly.\footnote{Work subsequent to [13] and [14] has extended this result in several directions. For our purpose, the original density theorem is enough.} 
The group $SL(2,{\bf Z}[i])$ has finite covolume in $SL(2, \bf C )$ and hence meets the conditions of the theorem, thereby enabling the taking over of the helicity content of any finite dimensional  irreducible representation of $SL(2,\bf C )$ as that of the (irreducible) representation of $SL(2,{\bf Z}[i])$ to which it restricts. 

An easy illustration of the density theorem is provided by the important special case of spin 1/2. The defining (left-chiral) representation $\rho_L$ of $SL(2,{\bf C}): \rho_L (\alpha)=\alpha \in SL(2,{\bf C})$ restricts to $SL(2,{\bf Z} [i])$ as $\rho_L (A)=A\in SL(2,{\bf Z} [i])$. Let $K$ be an operator on ${\bf C}^2$, the representation space of $\rho_L$, that commutes with $\rho_L (A)$ for all $A\in SL(2,{\bf Z} [i])$. Each of the Pauli matrices (multiplied by $i$) is in $SL(2,{\bf Z}[i])$ and so will commute with $K$ by assumption. Hence $K$ is a multiple of the unit operator and, by Schur's lemma, $\rho_L$ remains irreducible when restricted to $SL(2,{\bf Z} [i])$. The same conclusion -- and the same argument -- holds for the conjugate (right-chiral) representation. We may note incidentally that any two of the Pauli matrices can be picked to belong to the generators of $SL(2,{\bf Z}[i])$([15]).

The density theorem is critically important because of our general condition that only those representations are physically relevant whose `continuum limits' exist and are the Lie group representations we are used to. This is ensured if only those irreducible representations of $\hat{L}({\bf Z})$ are considered physical which are restrictions of (continuous) finite dimensional irreducible representations of $\hat{L}({\bf R})$.\footnote{It is useful to remember that even in the standard Wigner philosophy, certain URs of the Poincar\'{e} group are excluded from physics solely on the ground of lack of experimental support, e.g., those corresponding to imaginary mass or infinite/continuous spin. These criteria will be assumed to be valid in the discrete case as well. Thus URs of $\hat{P} ({\bf Z})$ which cannot be deboosted to rest -- corresponding to tachyons in continuum relativity -- are excluded.}

\section{Matter fields as unitary representations of $\hat{P}({\bf Z})$}

Having seen how the basic notions of momentum, mass and helicity carry over to a discrete space-time, we can turn now to our primary objective, that of identifying the elementary consistuents (particles) of matter with the fields that constitute Hilbert spaces of those irreducible URs of the discrete Poincar\'{e} group that are subject to the two physical criteria introduced in the last section. These are, to repeat, i) the orbit supporting a UR must be massless (which is an invariant property) or must have a real rest mass; and, ii) the representation of $\hat{L}(\bf Z )$ over which the UR ranges must be the restriction of a representation of $\hat{L}(\bf R )$. To these must be added a third, apparently more technical, criterion which arises as follows. 

A point $P$ of the Brillouin zone $B$ will be said to be a rational point if its coordinates $\{P_\mu\}$ are all rational multiples of $\pi$: $P_\mu =(q_\mu /d_\mu )\pi$, with $q_\mu, d_\mu \in \bf Z $, $-|d_\mu | \leq q_\mu \leq |d_\mu |$, and an irrational point otherwise. Every point in the orbit of $\hat{L}(\bf Z )$ through a rational (irrational) point is rational (irrational), since $\hat{L}(\bf Z )$ acts on $B$ by integral linear transformations followed by shifts by integral multiples of $\pi$. Consider rational orbits first. Reexpress the coordinates $\{q_\mu/d_\mu \}$ (dropping the factor $\pi$ for the time being) in terms of the lowest positive common multiple $D$ of \{$d_\mu\}$ , i.e., $P_\mu= Q_\mu/D$ with $-D\leq Q_\mu\leq D$ and no positive integer $D'<D$ exists such that $P_\mu=Q'_\mu /D'$ for any $\{Q'_\mu\}$ with $-D'\leq Q'_\mu \leq D'$; $\{Q_\mu/D\}$ will be referred to as the standard coordinates of the rational point $P$.  
 
Suppose now that the transform by $A$ of $P$ in standard form is not in standard form, i.e., $(AQ)_\mu$ has a common factor with $D$ for each $\mu$. Then the standard coordinates of $AP$ will have a denominator $D_A$ strictly less than $D$; if there is no common factor, the denominator of course remains unchanged. But the same argument applies also to $A^{-1}$ acting on $AP$, implying that $D$ cannot be greater than $D_A$. It follows that $\hat{L}(\bf Z )$ acts on every rational point without changing the denominator of  its standard coordinates. Since the numerators $\{Q_\mu \}$ lie between $-D$ and $D$, we can conclude that a rational orbit of $\hat{L} (\bf Z )$ in $B$ is a finite set. 

Recalling the identification of the orbit with the quotient of $\hat{L}(\bf Z )$ by the stabiliser, we see thus that the stabiliser of any rational point is a subgroup of finite index, in other words almost all of $\hat{L}(\bf Z ) $. The situation is not very different from that of the one-point orbit consisting of the origin $P=0$ -- the stabiliser has no finite dimensional UR from which to induce a UR of $\hat{P}(\bf Z )$ with a finite helicity spectrum (even when subjected to a finite set of subsidiary conditions, see the discussion of the massless URs of $\hat{P}(\bf R )$ in section 5). So our third  criterion is: rational orbits must be excluded from consideration in constructing physically reasonable URs of $\hat{P}(\bf Z )$. 

The next task is to determine the stabiliser of an irrational point $P$ for the action of $\hat{L}(\bf Z )$ on $B$. This group is 
$$   \Sigma_P: = \{A\in \hat{L}({\bf Z} ):(AP)_\mu = P_\mu \;\mathrm{mod}\; \bf Z \}. $$
It is most simply characterised in terms of the discrete Lorentz group $L(\bf Z )$ as the subgroup satisfying the conditions
$$   (\Lambda P)_\mu=P_\mu +N_\mu, \hspace{.5cm} \Lambda \in L(\bf Z ),     $$
for arbitrary integers $\{N_\mu \}$. For an irrational $P$, these conditions put strong restrictions on the admissible values of $N_\mu $: writing them as $(\Lambda-I)^{-1}N=P$ ($I$ is the unit matrix), we see that $P$ will be a rational point (since $(\Lambda-I)^{-1}$ has rational entries ($\Lambda -I$ is integral)) unless all $N_\mu$ vanish.
 
It follows that the stabiliser $\Sigma_P$ of an irrational $P$ for the action of $\hat{L} (\bf Z )$ on $B$ coincides with its stabiliser for the $\hat{L} (\bf Z )$ action on the whole of $M^*$; it is determined by the condition $(\Lambda P)_{\mu}=P_\mu$ exactly as in the continuum case. Thus, if $P$ is in a massive orbit (of rest mass $m$) of $\hat{L} (\bf Z )$, the corresponding stabiliser $\Sigma_m $ consists of all unitary matrices in $SL(2,{\bf Z} [i])$. It is a finite (of course) group of order 8, isomorphic to the quaternion group: $\Sigma_m =\{\pm 1, \pm i\tau_i \}$ where $\{\tau_i \}$ as earlier are the Pauli matrices.

Also as in continuum relativity, when $P$ is in a massless orbit, its stabiliser $\Sigma_0$ is the subgroup of $SL(2,{\bf Z}[i])$ consisting of upper triangular matrices (for a suitable choice of basis in $M^*$)
$$
s(\zeta,Z)=
\left (\begin{array}{cc}
\zeta & \zeta^{-1} Z \\
0 & \zeta^{-1}
\end{array} \right ); \; \zeta, \zeta^{-1}, Z \in {\bf Z}[i].
$$
The only elements of ${\bf Z}[i]$ with inverses in ${\bf Z}[i]$, namely the units, being $\zeta=\pm 1,\pm i$, the subgroup of diagonal matrices is the cyclic group ${\bf Z}_4$ and the subgroup of elements $s(1, Z)$ is the planar lattice of points $(\mathrm{Re}\;Z, \mathrm{Im}\;Z)$. Composition in $\Sigma_0$ is given by $s(\zeta_1,Z_1)s(\zeta_2,Z_2)=s(\zeta_1 \zeta_2, Z_1 +{\zeta_1}^2 Z_2)$, confirming that $\Sigma_0$ is indeed the discrete Euclidean group $E(2,{\bf Z})=SO(2,{\bf Z} )\vec{\times}{\bf Z}^2$, with $\zeta\in {\bf Z}_4$ acting on ${\bf Z}^2$ by multiplication by $\zeta^2$ -- which, as in the continuum case, is a reminder that $\hat{L} (\bf Z )$ covers $L(\bf Z )$ twice, thereby accommodating representations of half-integral helicities. 

We are thus confirmed in our expectation that the stabilisers are just the natural discretisations of their Lie group counterparts. The exercise also reinforces the choice of $SL(2,{\bf Z}[i])$ as the correct replacement for $SL(2, \bf C)$.

In constructing URs of $\hat{P}({\bf Z})$, whether massive or massless, we can now try to imitate Wigner's method in the continuum case, keeping in mind that all orbits are now discrete sets. Define momentum space fields as functions $\phi:O_B \rightarrow V$, where $O_B$ is $O_{B.m}$ or $O_{B,0}$ as the case may be and $V$ as before is the space of an irreducible representation $\rho$ of $\hat{L}(\bf R )$ (and hence, by the density theorem, of $\hat{L} (\bf Z )$). Given a set-section $\sigma:O_B \rightarrow \hat{L}({\bf Z})$, such fields form a Hilbert space ${\cal H}_{O_{B}}$ with scalar product
$$    \langle \phi, \psi \rangle =\sum_{P\in O_{B}}\langle \rho(\sigma(P)^{-1})\phi(P),\rho(\sigma(P)^{-1})\psi(P)\rangle _{V}      $$
(subject to the condition $\langle \phi ,\phi\rangle <\infty$). On this Hilbert space, we have a UR $U_{O_{B}}$ of $\hat{P}(\bf Z )$ given by 
$$  (U_{O_{B}}(A,X)\phi)(P)=\exp (iP_{\mu}X_{\mu})\phi (A^{-1} P), \: A\in \hat{L}({\bf Z} ), X\in T({\bf Z}).     $$
To repeat, all this is no more than a direct adaptation of the Wigner method to the discrete situation. It is to be noted in particular that the Fourier transform $\phi^*$ of $\phi$ is the field defined on space-time; since $\phi$ is a function on ${\bf T}^4$, $\phi^*$ is supported on ${\bf Z}^4$.  

But there are also some significant differences. The less serious one conceptually is that the discrete counterparts of the conventional field equations, which are Lie-algebraic statements to the effect that the stabiliser fixes points in an orbit (see section 5), have of course no infinitesimal version. Nor has the subsidiary condition which, in the massless case, enforces the requirement that admissible representations must have the vector subgroup of the Euclidean group as kernel. More seriously, while the orbit of $\hat{L}(\bf R )$ through $p\in M^*$ is the whole mass shell, massive or massless, containing $p$ -- i.e., $\hat{L}(\bf R )$ operates transitively on the submanifold of $M^*$ defined by a constant $p^2$ -- that is not the case for the action of $\hat{L}(\bf Z )$ on the torus mass shell; momenta which are linearly independent over the rationals cannot be connected by a discrete Lorentz transformation even if they are both massless or have the same rest mass. We can then take the orbit through a point $P'$ that is not in the orbit through $P$ (i.e., $P'$ and $P$ are rationally independent) but is in the same torus mass shell, and construct another irreducible UR of $\hat{P}(\bf Z )$, and so on. All of them will have the same stabiliser and the same rest mass and the same set of helicities; the Wignerian association of an elementary particle of a given mass and a given spin with an {\it irreducible} UR of $\hat{P} (\bf R )$ does not hold in discrete relativity.

There is then a choice to be made. One option is to associate each of the irreducible URs as constructed above with a particle type. Since all of them have the same rest mass and spin, distinguishing among them will require the introduction of new `quantum numbers' which, however, must still have their origin in the kinematics of discrete space-time itself. What might they be? It is unlikely that they can be related to any of the charges (`internal quantum numbers') of models currently in favour; among the more pragmatic reasons, we know of no infinite multiplets of particles of different charges all having the same mass and spin. 

An alternative possibility is that the different irreducible URs are superselection sectors. Recall that superselection rules are vetoes on the unrestricted superposability of states; they decompose the total state space into a collection of sectors whose direct sums do not represent states. An immediate physical consequence is that no observable can connect states from two distinct sectors (which in fact was the formulation they were first given [16]). The vetoes arise ([8,9]) from the non-additivity of inequivalent projective URs of a symmetry group and are determined by its 2nd cohomology group with suitable coefficients. In the present context, the symmetry group is $P(\bf Z )$ and its 2nd cohomology group is determined by that of its Lorentz subgroup $L({\bf Z})$. Group cohomolgy of discrete subgroups of Lie groups is a very active field but I have not been able to find the specific result needed here in the literature. (See also the discussion in section 3). Space-time symmetries can certainly give rise to superselection rules, the prime example being univalence, the rule that forbids the superposition of integral and half-integral spins, both in the continuum and, as we have seen in this paper, discrete situations. The question is whether in the discrete case the relevant cohomology group, which must contain the univalence  ${\bf Z}_2$, is in fact larger.\footnote{Very elementary examples of discretisations of Lie groups giving rise to additional nontrivial projective URs with interesting physical applications are provided by the symmetries of 2-dimensional spaces. Thus the cylinder group ${\bf S}^1\times{\bf R}$ (the natural group of an infinitely long strip with the two edges identified) has vanishing 2nd cohomolgy but its subgroup ${\bf S}^1\times {\bf Z}$ has nontrivial projective URs [17]. Similarly, the torus group ${\bf T^2}$ (acting on a rectangular space with pairs of opposite edges identified) has vanishing 2nd cohomology but its discrete subgroups ${\bf Z}_{n_1} \times{\bf Z}_{n_2}$ for positive $n_1$ and $n_2$ have non-zero cohomologies. The corresponding nontrivial projective URs are the underlying causes of phenomena related to quantum Hall effects ([18,19]).}    

If it turns out that there is no extra superselection structure a third option will be to give up the Wignerian criterion of associating a particle to an {\it irreducible} UR of $\hat{P} ({\bf Z})$. We are then free to assign a particle to a UR constructed, at least formally to start with, as the space of functions from the union of all irrational orbits of a given rest mass -- i.e., all of a given torus mass shell omitting rational orbits -- to a given irreducible representation of $\hat{L} (\bf R)$ (and hence, by the density theorem, of $\hat{L} ({\bf Z})$). Though highly reducible, such a UR of $\hat{P} ({\bf Z})$ will still be characterised by a unique rest mass and a unique spin. Transformations belonging to $\hat{L}(\bf R)$ but not to $\hat{L}({\bf Z})$ will connect different irreducible components of this UR which, in the continuum limit in some suitable sense, should tend to an irreducible UR of $\hat{P}(\bf R)$.  

Which of these options is the right one is a well-posed mathematical problem that is unsolved at present but solvable in principle. Until that is done, it seems prudent not to favour one over the others.  

\section{Physical effects: generalities}
The picture that has emerged can be summarised as follows. A discrete Minkowskian space-time incorporating a fundamental length $L$ -- to be thought of as one of the primordial constants of nature -- is fully capable of supporting the Wignerian correspondence between projective URs of its group of isometries and the quantum fields of elementary particles as defined by their masses and spins, subject to certain criteria of physical admissibility. Unsurprisingly, discretisation of special relativity makes no difference to the physics involving these particles as long as they are massless or are not subjected to large boosts taking them outside the conventional part of their mass shells. At energies and momenta resulting from large boosts there are new effects. This section and the next are devoted to a preliminary survey, mostly qualitative and occasionally speculative, of some such effects. (It is implicit that $L$ is (of the order of) the Planck length.) It is meant to be no more than an introduction to the issues involved and to convince the reader that the idea of a fundamentally discrete world is not to be rejected out of hand; it is premature to think of detailed quantitative computations.

Virtually all these deviations from conventional wisdom have their origin in the compactness of momentum space.\footnote{This is a very general feature, independent of any particular discretisation: the Pontryagin dual of a discrete abelian group is a compact abelian group.}
In general terms, its basic consequence is that energy and momentum are defined and conserved only modulo reciprocal lattice vectors, implying in turn that the notion of an invariant mass needs a reformulation.  The details are easiest to describe for the hypercubic discretisation employed here, for which the reciprocal lattice is also hypercubic, with the (Minkowskian) Brillouin zone given by (restoring its physical role to the lattice spacing $L$) $B=\{-\pi/L< P_\mu \leq\pi/L\}$.  Firstly, masslessness of a field/particle as an invariant concept survives discretisation: the torus light cone is mapped into itself under all boosts $P\rightarrow P'$. To remind ourselves, this property derives fundamentally from the absolute constancy of the speed of light, which itself is just the ratio of the spatial and temporal lattice spacings in conventional units (see section 3).\footnote{There is a subtle point of principle involved here. The general result, free from conventions, is that massless URs of $\hat{P} ({\bf Z} )$ exist and that the speed of propagation of the cirresponding particles, in particular the photon, is constant and equal to the ratio of the spatial and temporal lattice spacings. That our initial definition of the lattice using the speed of light  as a conversion factor guarantees this outcome is far from obvious at the outset. A reassessment of the fundamental significance of the speed of light, both logically and historically, will be found in a review now in preparation.}
Next, though a non-zero mass remains invariant under `small' boosts $p\rightarrow p'$ with $p,p'\in B$, that is not the case for `large' boosts taking $p$ out of $B$, resulting in discontinuous changes in $P$ at the boundary of $B$. The physically meaningful (and useful) notions are that of a rest mass and of a mass shell of rest mass $m$, the orbit of the Lorentz group in $B$ through the point $(m,0,0,0)$ (the torus mass shell of section 7). Consequently, when a particle of rest mass $m$ undergoes large boosts, it traverses intermittently regions of the mass shell where its speed becomes superluminal. In the limit of an infinitely large boost, the speed will tend to the speed of light, passing through, in the process, an infinite number of superluminal phases (Figure 1). It is also clear that the momentum intervals in which the particles have a transluminal speed grow as its rest mass $m$ increases (reminder: Figure 1 is for $m$ of the order of $L^{-1}$). 

In looking at the observable effects of these unfamiliar kinematical properties, a reservation will be in order: our inability, as of now, to resolve the question of reducibility/degeneracy of URs of $\hat{P}(\bf Z )$, discussed at the end of section 7. As long as the identity of an elementary particle is primarily decided by its mass and spin -- i.e., assuming, as seems justified in our present state of knowledge, that internal charges and the concomitant gauge interactions have an origin outside space-time geometry -- irreducibility by itself may not be a critically important criterion. Such a position would require us to accept that matter considered elementary at currently accessible scales, including the invisible quarks and gluons, will continue to be describable as elementary quantum fields of the discrete Poincar\'{e} group. That would be a big extrapolation; on the natural scale of the Planck mass, the known elementary particles are massless to an extraordinarily good approximation, to an accuracy of some eighteen orders of magnitude. Aside from the cosmological constant problem, there are no other examples in fundamental physics of such an enormous deviation from naturalness. One can wonder whether there actually are -- or were -- elementary particles of mass comparable to $L^{-1}$ but, as of now, that would be idle speculation. So,
in checking the exotic effects of discreteness against facts on the ground -- or, rather, in the heavens -- no specific assumption will be made here regarding the possible presence of such ultra-heavy matter particles through the history of the universe. (Modulo poorly understood phenomena such as dark matter or conjectural entities like the inflaton, modern cosmological theory, including that of the very early universe, has no place for any exotic particles). 

Of the direct effects of discreteness, testing for deviations from homogeneity and isotropy of space is no more or less than determining its `crystal structure'. Intuitively it is clear that any such procedure will need to measure spatial (and, in the case of homogeneity of time, temporal) resolutions finer than $L$ (the Planck length by default). For illustration, for a beam of particles (emitted somewhere in the universe) interacting with the fields $\phi^*$ concentrated at the `lattice sites' of space to produce a measurable (macroscopic) diffraction pattern, the wavelength of the particle has to be of the order of the lattice spacing, just from Bragg's law. There is no need to add that such tests are out of reach by very many orders of magnitude. 

An essentially similar conclusion holds for tests of isotropy, even though angles (as opposed to lengths) are not restricted to be discrete. For a periodic lattice such as the cubic one considered here, a reasonable statement  of isotropy would appear to be that an arbitrary straight line through any chosen origin 0 (`the line of sight') should pass through an infinite number of lattice points $lX$, $l=0,1,2,\cdots; \; X\in M(\bf Z )$. (It is enough that it passes through one such point, say $l=1$). This sharp formulation however does not quite work. When the line of sight through 0 is, for example, in the 1-2 plane: $X=(0,X_1 ,X_2 ,0)$, and makes an angle $\theta$ with the 1 axis, $\tan\theta=X_2 /X_1$ will be rational whereas there is a general result (part of what is known as Niven's theorem [20]) which says that the only rational values of the tangent function at rational angles (rational multiples of $\pi$) are $0, \pm 1$. In other words, if the line of sight is at a rational angle (except for $0$ and $\pi/4$ in the first quadrant) it will pass through no lattice points at all. The remedy suggests itself: the continuity of the tangent function implies that, given a point $X$, there are angles $\theta$ such that the corresponding lines of sight pass as close as desired to the points $lX$ for all $l$ less than any given finite integer. This slightly weaker formulation of isotropy is perfectly satisfactory both theoretically and observationally. 

\section{Physical effects: some qualitative cosmology}

The kinematical effects of the boundedness of momentum and energy seem, likewise, to be far beyond the capabilities of any controllable experiment even in a remote future: how to boost particles to energy/momentum of magnitudes of the order of $10^{20}$ GeV, i.e., to the edge of the Brillouin zone? That leaves cosmic observations as the only realistic test of discreteness, especially those which are sensitive to the physics of processes which we believe happen in the early universe. One such is the deboosting\footnote{Throughout this article, I have avoided using the terms `acceleration' and `deceleration' in favour of `boost' and `deboost' for the reason that energy and momentum are always within $B$ and change discontinuously when they hit its boundary. Velocity, defined as $\vec{V}=\vec{P}/P_0$ so as to be consistent with the continuum limit, also has discontinuities. Boost operations are well-defined in $B$ and, along any of the three spatial axes, can be ordered by the value of the parameter (sometimes called rapidity) they depend on. The essential point is that an infinite boost does not result in infinite energy, momentum or velocity. For the same reason, the evolution of the discrete universe from the big bang to the present is to be thought of as parametrised by a monotonically decreasing (average) boost parameter of matter particles rather than energy or velocity.}
of matter in the initial hot dense `stuff' that leads to expansion and cooling (in the accepted, `standard', cosmological model). In the discrete world the process will go through a series of steps in which every particle of matter crosses the light barrier, into and out of the light cone, alternately (Figure 1 with the arrows reversed shows the final stages of the process) until it makes its final transit into the interior (to the conventional part of its mass shell), after which it will behave exactly as in conventional special relativity. The question is how these repeated transluminal episodes fit in with the generally agreed post-big-bang picture of the early cosmos. 

Considering that all transits including the final one into the conventional regime, the subluminal world of our experience, took place when particle energies were close to $L^{-1}$ (see Figure 1 and recall again that what decreases monotonically with time as the universe evolves is not energy but the boost parameter) and therefore in the very remote past about which we know little, the answer to the question has to be that cosmology as we know it today is definitely not in conflict with the above picture. Indeed, any possibility of identifying superluminal expansion originating in discreteness is likely to be subsumed in the by-now standard -- though still mysterious -- cosmic inflation. The two differ in their origins. The superluminal propagation arises here kinematically and in a purely special relativistic context -- gravity plays no role -- whereas the root cause of inflation remains unsettled despite the enormous amount of work,  mixing the usual quantum field theory ideas with general relativity, that has gone into it. The details of the process are also different: episodic with a long series of superluminal phases in our case and one big blowout in inflation. One may then ask: if the world is really discrete, can one do without inflation in its current formulation(s)? In partial answer, let us note the encouraging fact that  the energy of a massive particle vanishes -- the torus mass shell intersects the cube that is the subspace $P_0 =0$ of the Brillouin zone (the horizontal axis in Figure 1 is its projection) -- in a repeated series of events until it makes its final transit into the conventional regime. Such events occur during highly superluminal phases: when $P_0 =0$, the velocity is infinite.\footnote{The sequential superluminal expansion implied by discreteness and advocated here has nothing conceptually in common with models, some of them very explicit (an early example is [21]), which postulate that the speed of light itself has changed enormously over cosmic time scales, cause unknown. In the discrete universe the agency of homogenisation is not radiation (massless particles have an absolutely constant, conventional, speed) but massive matter. I add that cosmological issues related to inflation continue to see an enormous amount of activity, as regards both its conceptual moorings and their empirical validation; for a thorough review of the current situation, see [22].} 

Aside from global consequences like superluminal expansion, the lack of strict energy-momentum conservation -- conservation only modulo $2\pi/L$ -- also leads directly to a class of Planck-scale Umklapp processes, namely interactions of individual elementary particles in which some of them in the initial or/and final states would have energies or momenta or both outside the Brillouin zone if they were strictly conserved, i.e., before enforcing the periodicity of reciprocal space. Umklapp phenomena in crystal physics\footnote{An excellent general reference is the book of Peierls ([23]) who first recognised the phenomenon and its origin in the discrete symmetry groups of crystals.} 
concern only the momenta and velocities of the relevant elementary excitations (mainly phonons and electrons) -- energy is always strictly conserved since time is not discrete -- and they influence chiefly the transport properties in crystalline media. Moreover, the medium itself is not discrete -- only the symmetry group is -- and is of finite size. Consequently, the effect of periodicity is just to modulate the 1-particle wave function (Bloch's theorem) and the apparent violation of momentum conservation is only an artefact of translating all momenta back to the Brillouin zone. Physically, the momentum of the total system consisting of the elementary excitations and the crystal is conserved; the crystal as a whole recoils to balance the loss or gain of momentum by the crystalline equivalent of matter.

In contrast, when the `crystal' is space-time itself, the violation of energy-momentum conservation must be regarded as a real physical phenomenon: there is no sensible physical meaning that can be given to the notion of Minkowskian space-time having kinematical attributes like energy and momentum. Nevertheless, since very few processes in solids are measurably sensitive to crystal recoil,\footnote{Exceptions would be resonance absorption of photons and neutrinos.}
we can adapt the methodology of crystal physics -- extending them to include energy in addition to momentum -- for a first orientation on recoilless space-time Umklapp interactions. An interaction in which the initial state has particles each of which is in its conventional regime, with momenta $\{P_\mu ^{(i)} \}$, can result in two general classes of final states: either i) all of the particles in the final state, with momenta $\{ P_{\mu}^{'(j)} \}$, are in their conventional regimes -- normal processes in the language of crystal physics; or ii) some of the particles will go outside the Brillouin zone $B$ if energy and momentum were strictly conserved, $\sum_i P_\mu ^{(i)}=\sum_j P_{\mu}^{'(j)}$, which have then to be translated back to $B$ -- Umklapp processes. The general conservation law, always valid, is    
$$   \sum_i P_\mu ^{(i)}-\sum_j P_{\mu}^{'(j)}=\frac{2\pi N_\mu}{L}    $$
for some $N_\mu \in \bf Z$, depending on the values of $P_\mu ^{(i)}$ and $P_\mu ^{'(j)}$.
 
As the simplest possible example of the new kinematics, consider first the decay at rest of a particle of rest mass $M$ into two particles of equal rest mass $m$. Assuming strict momentum conservation, the daughter particles will have momenta in opposite directions, say along the 1-axis, of equal magnitude $k$. Assuming also strict energy conservation, we have $4(k^2 +m^2 ) =M^2$ or $k=\sqrt{M^2 /4-m^2 }$. From the reality of $k$ (for the decay to be kinematically possible), $m<M/2$. And since $M<\pi/L$ (from the definition of the rest mass), it follows further that $k<M/2< \pi/2L$. No energy or momentum is outside the Brillouin zone and the decay is a normal process. No surprise here.

But this apparently trivial example becomes interesting when we consider its inverse reaction: the collision of two equal rest mass particles to produce a final state whose centre-of-mass is at rest, including the possibility of a single particle at rest. Assuming strict conservation, the total energy of the final state, $E=2\sqrt{k^2 +m^2 }$ (which is also its invariant mass), will exceed the Brillouin zone bound $\pi/L$ if either $k>\pi/2L$ or $m>\pi/2L$ (or both). For such values of $k$ and $m$ strict conservation cannot hold: an Umklapp shift down has to be made (one shift is enough since $k$ and $m$ are bounded above by $\pi/L$), resulting in a final state energy of $2\sqrt{k^2 +m^2 }-2\pi/L$. 

The example can be extended to more general kinematic situations in a straightforward manner though the energy-momentum book-keeping can get quite involved. The inescapable fact is that dramatic degradations of energy and momentum in interactions of particles of Planck scale masses and energies are a generic consequence of discrete special relativity. Once again, the only realistic possibilities of testing the effect would seem to involve the physics of the very early cosmos but
the fact that the shift can result in relatively moderate final energies gives some hope. Perhaps ultra high energy cosmic rays (with energies some 8 or 9 orders of magnitude smaller than $1/L$) are the final result of such degradation, tantalising relics from the early history of the universe, rather than the product of some unknown cosmic mechanism involving extreme acceleration. 

While on the subject of possible future lines of work, we should also note that the breakdown of conventional energy momentum conservation -- but in a precisely quantifiable way -- has other consequence which are relevant to the study of the early universe, specifically in its hot dense phase. In general terms, what is required is a reformulation of thermodynamics in extreme conditions in which the kinematics of binary collisions is governed by the new conservation laws. That is a well-posed problem but, obviously, a very challenging one. We can surely expect a reliable theory incorporating them to lead to deviations from currently popular cosmological models of the young hot dense universe. Whether they will turn out to be theoretically and observationally acceptable is for the future to decide. What is generally gratifying in the meantime is that the knowledge we already have accommodates the effects of discreteness in the structure of space-time quite comfortably without any extra assumptions. Theoretically, the deep connection between the existence and properties of elementary particles and the symmetries of space-time survives intact and, observationally, none of our hard-won insights into the mechanisms of the physical world is put at risk. One can then hope that the hypothesis of a fundamental length may, through the novel kinematics it entails, provide an opening into some of the more ad hoc features of current cosmological theory and into fundamentally kinematic phenomena for which there is no convincing basis yet (such as, obviously, dark matter and dark energy). To conclude, it is worth repeating that these novel effects do not invoke the curvature of space-time in any way; the consequences of the gravitational aspects of general relativity will be above and beyond them. 

\vspace{1cm}
It seems fitting to end with Riemann's much-cited exhortation on the need to respect both experience and logic in studying the nature of space (in his famous Habilitation lecture ``On the Hypotheses which Lie at the Foundation of Geometry'', 1854):
\begin{quote}
 . . . it is a necessary consequence that . . . those properties which distinguish space from other conceivable triply extended (3-dimensional) quantities can only be deduced from experience. Thus arises the problem of seeking out the simplest data from which the metric relations of space can be determined . . . These data, like all data, are not logically necessary, they are hypotheses; one can therefore investigate their likelihood, . . . and afterwards decide on the legitimacy of extending them beyond the bounds of observation, both in the direction of the immeasurably large {\it and} (my italics) in the direction of the immeasurably small . . .
\end{quote}

\vspace{1cm}

I acknowledge with warm thanks the invaluable advice of M. S. Narasimhan, M. S. Raghunathan, Parameswaran Sankaran and T. N. Venkataramana on the mathematical aspects of this work, first addressed, in a different context, in an unpublished manuscript ([24]). On questions related to the current state of knowledge of early cosmology, especially of the role of inflation, I have benefited greatly from the explanations of Swagat Mishra. For that, and for his help with Figure 1, I express my gratitude. Finally, I acknowledge here the hospitality of the Inter-University Centre for Astronomy and Astrophysics (Pune) over the years that helped bring this work to a (tentative) close. 

\vspace{.4cm}

{\bf Bibliography}

\vspace{.2cm}

[1] S. Hossenfelder, Living Rev. Relativity {\bf 16}, 2 (2013).

[2] R. Loll, arXiv:gr-qc/9805049 (1998).

[3] R. Williams, J. Phys. Conference Series {\bf 33}, 38 (2006).

[4] E. P. Wigner, Ann. Math. {\bf 40}, 149 (1939). 

[5] V. Bargmann and E. P. Wigner, PNAS {\bf 34}, 211 (1048).

[6] Wigner, {\it Group Theory and its Applications to the Quqntum Mechanics of Atomic Spectra} (Academic Press, New York, 1959). 

[7] M. S. Raghunathan, Rev. Math. Phys. {\bf 6}, 207 (1994).

[8] P. P. Divakaran, Rev, Math, Phys. {\bf 6}, 167 (1994).

[9] P. P. Divakaran, Phys. Rev. Letters {\bf 79}, 2159 (1997).

[10] R. F. Streater and A. S. Wightman {\it PCT, Spin-Statistics and all that} (Benjamin, New York, 1964).

[11] G. W. Mackey, Ann. Math. {\bf 55}, 101 (1952); {\bf 58}, 143 (1953).

[12] M. Sugiura, {\it Unitary Representations and Harmonic Analysis} (2nd ed.) (North Holland/Kodansha, Amsterdam/Tokyo, 1990).
 
[13] A. Borel, Ann. Math. {\bf 72}, 179 (1960).

[14] M. S. Raghunathan, {\it Discrete Subgroups of Lie Groups} (Springer, Berlin, 1972).

[15] R. G. Swan, Adv. Math. {\bf 6}, 1 (1971).

[16] G-C Wick, A. S. Wightman and E. P. Wigner, Phys. Rev. {\bf 88}, 101 (1954).

[17] G. Date and P. P. Divakaran, Ann. Phys. {\bf 309}, 429 (2004).

[18] P. P. Divakaran and A. K. Rajagopal, Int. J. Mod. Phys. B {\bf 9}, 261 (1995).    .

[19] P. P. Divakaran, unpublished.

[20] I. Niven, {\it Irrational Numbers} (Mathematical Association of America, New York, 1956).

[21] J W Moffat, Int. J. Mod. Phys. D {\bf 2}, 351 (1993).

[22] Swagat S. Mishra. ``Cosmic Inflation: Background dynamics, Quantum fluctuations and Reheating'', arXiv: 2403.10606[gr-qc] (2024). 

[23] R E Peierls, {\it Quantum Theory of Solids} (Clarendon Press, Oxford, 1955). 

[24] P. P. Divakaran, arXiv:hep-lat/0204027 (2002).

\end{document}